\documentclass{emulateapj}
\usepackage{apjfonts}
\usepackage{epsf}
\begin{document}

\slugcomment{Accepted for ApJL}
\shortauthors{J. M. Miller et al.}
\shorttitle{Winds or Jets}

\title{The Disk-Wind-Jet Connection in the Black Hole H 1743$-$322}

\author{J.~M.~Miller\altaffilmark{1},
        J.~Raymond\altaffilmark{2},
        A.~C.~Fabian\altaffilmark{3},
        C.~S.~Reynolds\altaffilmark{4},
        A.~L.~King\altaffilmark{1},
        T.~R.~Kallman\altaffilmark{5},
        E.~M.~Cackett\altaffilmark{6},
        M.~van der Klis\altaffilmark{7},
        D.~T.~H.~Steeghs\altaffilmark{8}}

\altaffiltext{1}{Department of Astronomy, University of Michigan, 500
Church Street, Ann Arbor, MI 48109-1042, jonmm@umich.edu}

\altaffiltext{2}{Smithsonian Astrophysical Observatory, 60 Garden
  Street, Cambridge, MA, 02138}

\altaffiltext{3}{Institute of Astronomy, University of Cambridge,
  Madingley Road, Cambridge, CB3 OHA, UK}

\altaffiltext{4}{Department of Astromoy, University of Maryland,
  College Park, MD, 20742, USA}

\altaffiltext{5}{Laboratory for High Energy Astrophysics, NASA Goddard
  Space Flight Center, Code 662, Greenbelt, MD 20771}

\altaffiltext{6}{Department of Physics and Astronomy, Wayne State
  University, 666 West Hancock Street, Detroit, MI 48201}

\altaffiltext{7}{Astronomical Institute ``Anton Pannekoek'',
  University of Amsterdam, Science Park 904, 1098XH, Amsterdam, NL}

\altaffiltext{8}{Department of Physics, University of Warwick, Coventry CV4 7AL, UK}


\label{firstpage}

\begin{abstract}
X-ray disk winds are detected in spectrally soft, disk--dominated
phases of stellar-mass black hole outbursts.  In contrast, compact,
steady, relativistic jets are detected in spectrally hard states that
are dominated by non-thermal X-ray emission.  Although these
distinctive outflows appear to be almost mutually exclusive, it is
possible that a disk wind persists in hard states but cannot be
detected via X-ray absorption lines owing to very high ionization.
Here, we present an analysis of a deep, 60~ksec {\it Chandra}/HETGS
observation of the black hole candidate H~1743$-$322 in the low/hard
state.  The spectrum shows no evidence of a disk wind, with tight
limits, and within the range of ionizing flux levels that were
measured in prior {\it Chandra} observations wherein a wind was
clearly detected.  In H 1743$-$322, at least, disk winds are actually
diminished in the low/hard state, and disk winds and jets are likely
state-dependent and anti-correlated.  These results suggest that
although the launching radii of winds and jets may differ by orders of
magnitude, they may both be tied to a fundamental property of the
inner accretion flow, such as the mass accretion rate and/or the
magnetic field topology of the disk.  We discuss these results in the
context of disk winds and jets in other stellar-mass black holes, and
possible launching mechanisms for black hole outflows.
\end{abstract}

\section{Introduction}
Recent {\it Chandra}/HETGS observations of stellar-mass black holes
and neutron stars have revealed disk winds through blue-shifted X-ray
absorption lines (Miller et al.\ 2006a,b; Kubota et al.\ 2007; Miller
et al.\ 2008; Ueda, Yamaoka, \& Remillard 2009; Neilsen \& Lee 2009,
King et al.\ 2012; Ponti et al.\ 2012, King et al.\ 2012b).  Such
winds are not a negligible part of the accretion flow: estimates for
the mass outflow rate in winds range from a fraction of the accretion
rate through the disk, to many times greater than the mass accretion
rate ($\dot{m}$) through the disk.  A full understanding of disk
accretion now requires an understanding of such outflows, and how they
are driven.  More broadly, these winds may provide an important
grounding for tentative evidence of ionized winds from the inner disk
of AGN (Tombesi et al.\ 2012; King, Miller, \& Raymond 2012, King et
al.\ 2012b).

Stellar-mass black hole disk winds appear to be state--dependent: they
are detected in spectrally--soft, disk--dominated states, but are
not clearly detected in the ``low/hard'' state (Miller et al.\ 2006a,
2008; Neilsen \& Lee 2009, Blum et al.\ 2010; Ponti et al.\ 2012, King
et al.\ 2012).  In contrast, relativistic radio jets are ubiquitous in
the ``low/hard'' state (Fender, Belloni, \& Gallo 2004), but quenched
in disk--dominated soft states (e.g. Russell et al.\ 2011).  Thus, it
is possible that winds and jets are anti--correlated, but related.

An anti-correlation might offer rare clues to the mechanisms
that drive wind and jets.  However, the nature of changes to the
accretion inflow geometry and radiative processes across state
transitions remains unclear (e.g. Esin, McClintock, \& Narayan 1997;
Reis, Fabian, \& Miller 2010).  Even the apparent absence of
disk winds in the ``low/hard'' state may only be an observational
effect driven by over-ionization of the gas.  A wind might continue
unabated, but simply be impossible to detect through absorption
lines owing to a higher ionizing photon flux level.

Ponti et al. (2012) suggested that a higher ionizing flux could not
explain the lack of observable wind features in the hard state.
However, they did not consider whether the associated change in
spectral shape could provide the required change in ionization.
Nielsen and Homan (2012) carried out a more complete analysis to show
that the extremely dense wind seen in the soft state of GRO J1655-40
by Miller et al. (2006b) could not also be present in a harder state
(likely an ``intermediate'' state) observed a few days earlier, with
the difference in absorption lines explained by ionization alone.
That extreme wind state has only been reported in one other BHB
observation, and the driving mechanism may be different from that of
the more normal winds in which only Fe XXVI and perhaps Fe XXV are
detectable.

In this paper, we compare the Fe absorption lines seen in the
soft state of H1743-322 with the lack of absorption lines in a true
low/hard state.  We find that the difference in photoionization rate
cannot account for the spectra, but that the wind must be genuinely
suppressed.  We use a quantitative approach to the difference in
absorption line strengths to reach a better understanding of disk
winds based on analysis of a deep Chandra HETGS observation of H
1743-322 in the low/hard state, focusing on the physical parameters
implied by the apparent absence of the wind.

\begin{figure}
\includegraphics[scale=0.35,angle=-90]{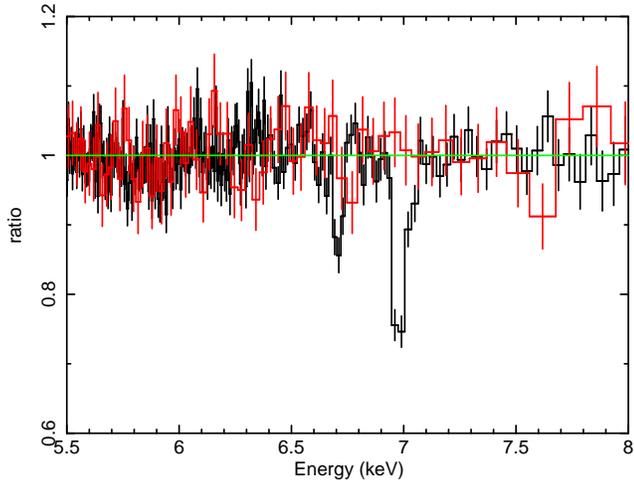}
\figcaption[t]{\footnotesize The figure above shows the line spectra
  from two {\it Chandra}/HETGS observation of H 1743$-$322.  The data
  have been divided by a simple continuum and binned for visual
  clarity.  The observation in black was obtained in a disk--dominated
  phase; it is listed as ``Observation 1'' in Miller et al.\ (2006a).
  A disk wind was detected through the blue-shifted Fe XXV and Fe XXVI
  absorption lines in that specturm.  The deep low/hard state
  observation considered in this paper is shown in red.  No
  significant absorption lines are evident, and restrictive upper
  limits are obtained through direct fits.}
\end{figure}
\medskip

\section{Observations and Data Reduction}
During its 2003 outburst, H 1743$-$322 was observed simultaneously
using the {\it Chandra}/HETGS and {\it RXTE} on four occasions (Miller et al.\ 2006a).  All four of those observations captured relatively soft
flux states.  Three could be roughly classified as ``high/soft''
states.  The remaining observation (second in the sequence) likely
captured a harder (but still luminous) ``very high'' state.

H 1743$-$322 was again observed with the {\it Chandra}/HETGS and RXTE
in 2010.  The {\it Chandra} observation started on 2010 August 08 at
23:03:48 UT, and was 60.5~ksec in duration.  RXTE observed H
1743$-$322 simultaneously with {\it Chandra}, starting on 2010 August
09 at 05:35:51 UT for a total duration of 6.1 ksec.  Rapid analysis of
the RXTE observation confirmed that H 1743$-$322 was in a
``low/hard'' state at the time of these observations (Belloni et al.\ 2010).

The {\it Chandra} High Energy Transmission Gratings (HETG) were used
to disperse the incident flux onto the Advanced CCD Imaging
Spectrometer ``spectroscopic array'' (ACIS-S).  To prevent photon
pile-up, the ACIS-S array was operated in continuous clocking or
``GRADED\_CC'' mode, which reduced the frame time to
2.85 msec.  For a discussion of this mode, please see Miller et
al.\ (2006a).  All {\it Chandra} data reduction was accomplished using
CIAO version 4.4.  Time-averaged first-order HEG and MEG spectra were
extracted from the Level-2 event file.  Redistribution matrix files
(rmfs) were generated using the tool ``mkgrmf''; ancillary response
files (arfs) were generated using ``mkgarf''.  The first-order HEG
spectra and responses were combined using ``add\_grating\_orders''.
The spectra were grouped to require a minimum of 10 counts per bin.

The standard RXTE pipeline spectral files and responses for the PCA
and HEXTE were downloaded from the HEASARC archive, and employed for
spectral fitting.  A systematic error of 0.6\% was added to the PCA
spectrum in quadrature.  All spectral analyses were conducted using
XSPEC version 12.6.0.  All errors quoted in this paper are 1$\sigma$
errors.

\section{Analysis and Results}
The central questions in this paper require estimates of the ionizing
flux and column density in each observation of H 1743$-$322.  This can
be done through photoionization modeling, but a simpler and more
direct approach is to measure the equivalent width of absorption lines
since ${\rm EW} \propto {\rm N}$ when the absorbing gas is on the
linear part of the curve of growth.

\subsection{The Spectral Continuum in the 2010 Low/Hard State}
We fit the combined first-order {\it Chandra}/HEG and RXTE/PCA spectra
of H~1743$-$322 jointly.  The HEG spectrum was fit was limited to the
1.2--9.0~keV band, owing to calibration uncertainties and poor signal
on either side of this range.  The PCA spectrum was fit in the
3.0--30.0~keV band, again owing to calibration uncertainties on either
side.  In the joint fits, a simple constant was allowed to float
between the spectra to account for differences in the flux calibrations.

A fit with a simple absorbed power-law model with $\Gamma = 1.77\pm
0.01$ does not give a formally acceptable fit ($\chi^{2}/\nu = 1.56$),
but it does characterize the flux well.  It is likely that the poor
fit is driven by uncertainties in the cross-calibration of the
instruments.  When each spectrum is permitted to derive its own
power-law index in a joint fit, a value of $\Gamma = 1.93\pm 0.01$ is
found for the HEG spectrum while $\Gamma = 1.71\pm 0.01$ is found for
the RXTE/PCA spectrum, and a much better fit is derived ($\chi^{2}/\nu
= 1.13$).  If the steeper index is assumed to be the right
time-averaged value for the lengthy {\it Chandra} observation, it
leads to a low value for the 8.8--30~keV ionizing flux (see below).
To be conservative, then, we simply adopt the power-law index derived
in the joint fit ($\Gamma = 1.77$) as an approximate value, and derive
the unabsorbed 8.8--30~keV flux based on that model.

\subsection{Limits on Absorption Lines in the 2010 Hard State}
To test for the presence of Fe XXV and Fe XXVI absorption lines in the
low/hard state spectrum of H 1743$-$322, we added Gaussians
at 1.850~\AA~ and 1.780~\AA~ (6.700~keV and 6.970~keV, respectively;
Verner, Verner, \& Ferland 1996).  The range of line widths and
velocity shifts measured in the line detections reported in Miller et
al.\ (2006a) were sampled in order to ensure consistency and
conservative limits.  We measure 90\% confidence
upper limits of ${\rm EW} \leq 0.58$~ m\AA~ and ${\rm EW} \leq 0.42$~
m\AA~ for Fe XXV and Fe XXVI, respectively.  These limits are a factor
of several lower than the line detections reported in Miller et
al.\ (2006a).  Figure 1 shows data/model ratio spectra from a prior
{\it Chandra} observation and the low/hard state considered here.

Larger potential velocity shifts were also examined, since one means
of increasing the ionization of a gas is to accelerate it.  A
potential feature is apparent at 1.834~\AA~ (6.760 keV).  However,
this feature is merely noise: first, the feature is not significant at
even the 2$\sigma$ level; second, it is unlikely that Fe XXV would be
observed in the absence of Fe XXVI (see, e.g., Kallman \& McCray
1982), especially if the gas is potentially {\it more} highly ionized
than when both were detected.  There is no evidence for an Fe XXVI
line at a velocity shift required if the feature at 6.760~keV line is
identified with Fe XXV.

Limits on the line equivalent widths obtained with {\it Chandra}
(again, 0.58~m\AA~ and 0.42~m\AA, for Fe XXV and Fe XXVI,
respectively), are tighter than those obtained in a prior {\it Suzaku}
observation in the low/hard state (0.97~m\AA~ and 0.64~m\AA~, Blum et
al.\ 2010).  Importantly, whereas the ionizing flux derived from the
{\it Suzaku} observation was {\it higher} than during the prior
observations wherein a disk wind was detected, the ionizing flux
derived in the new {\it Chandra} observation is {\it lower} than two
of the three prior observations with detections.

\begin{figure}
\includegraphics[scale=0.50]{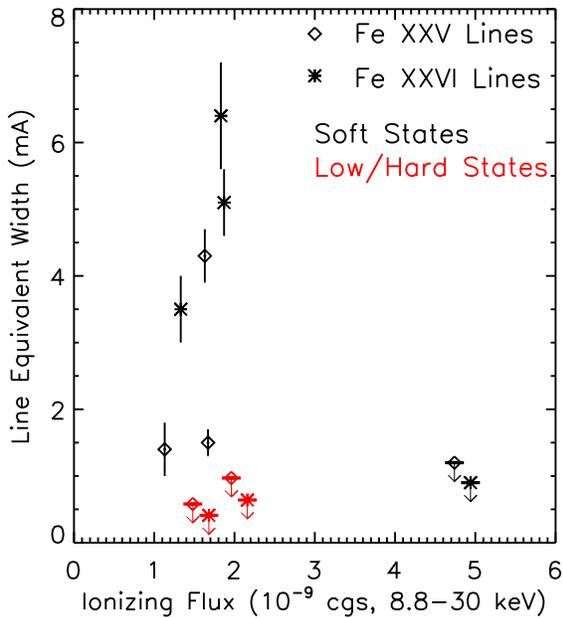}
\figcaption[t]{\footnotesize Blue-shifted Fe XXV and Fe XXVI lines are
  the primary tracers of disk winds in stellar-mass black hole
  spectra.  The figure above plots line equivalent width (a proxy for
  the column density) versus the ionizing flux, based on Chandra
  observations of H~1743$-$322.  Points corresponding to Fe XXVI were
  artificially shifted upward in ionizing flux by 0.2 for visual
  clarity.  Upper limits (90\% confidence) are characterized with
  arrows.  Black points represent measurements obtained in soft,
  disk--dominated states while red points signify measurements made in
  the low/hard state.  Line limits from a {\it Suzaku} low/hard state
  observation (Blum et al.\ 2010) are also included (the higher flux
  pairing of points).}
\end{figure}
\medskip

\subsection{Line Equivalent Widths Versus Ionizing Flux}

In order to ionize He-like Fe XXV, photons with ${\rm E} \geq 8.8$~keV
are required (E $\geq 9.3$~keV is required for Fe XXVI; Verner,
Verner, \& Ferland 1996).  To characterize the ionizing flux in each
observation of H 1743$-$322, then, we derived the (unabsorbed) flux in
the 8.8--30.0~keV band, based on direct fits to the spectral
continuum.  For plausible spectral indices, the 30--100~keV photon
flux represents $\leq 1$\% of the photoionization rate.  Given that
some low/hard state spectra have a spectral break at approximately
30~keV, and given that photons above 30~keV are unimportant, the
8.8--30.0~keV band is a meaningful and pragmatic range for
characterizing the ionizing flux.

Spectral fits to {\it Chandra} and {\it RXTE} observations in bright
states are described in Miller et al.\ (2006a).  These models were
used to calculate the unabsorbed 8.8--30~keV flux for each bright
phase observation.  Where multiple RXTE observations were made within
a single {\it Chandra} exposure, their ionizing flux values
were averaged.  The same procedure was used to calculate the ionizing
flux in the {\it Suzaku} low/hard state observation (Blum
et al.\ 2010).  For the {\it Chandra} low/hard state observation that
is the focus of this paper, the unabsorbed 8.8--30~keV flux was
calculated based on the fits in Section 3.1.

Figure 2 plots line equivalent widths versus the 8.8--30.0~keV
ionizing flux, based on {\it Chandra} detections and limits.  The {\it
  Chandra} upper limits in the ``low/hard'' state reported in this
work are lower than the ionizing flux levels in two of three prior
cases wherein a disk wind was detected (Miller et al.\ 2006a).  It is
probable, then, that the wind is diminished or even quenched
in the ``low/hard'' state.  

The strongest Fe XXVI line detected in the 2003 outburst corresponded
to $N_{\rm XXVI} = 5.5\pm 0.7 \times 10^{17}~ {\rm cm}^{-2}$, or
$N_{\rm H} = 1.8 \times 10^{23}~ {\rm cm}^{-2}$, after accounting for
the abundance of Fe ($3\times 10^{-5}$ is typical of the literature)
and ionization fraction (Fe XXVI $/$ Fe XXVII $\leq$ 0.1, e.g. Kallman
\& McCray 1982).  In simple terms, then, the new observation implies
$N_{\rm H} \leq 0.86 \times 10^{22}~ {\rm cm}^{-2}$.  In the high
ionization limit, e.g. when Fe XXV is not seen, $f_{XXVI} \propto n
q_{rec} / L_{X} r^{2}$ (where $f_{XXVI}$ is the ionization fraction of
Fe XXVI, $n$ is the number density, and $q_{rec}$ is the recombination
rate coefficitient).  Thus, if $r$ stays the same, $f \propto n$.
Therefore, reducing the density by a factor of $\sim$3 would reduce
the total column by a factor of $\sim3$, and the corresponding
reduction in $\xi$ would reduce the ionization fraction (Fe XXVI$/$Fe)
by $\sim$3, in combination giving an order of magnitude reduction in
$N_{XXVI}$ (the observed quantity).  This would require
$\dot{M}_{wind}$ to be at least a factor of $\sim$3 lower.  Note that
the strongest line detected in 2003 did not show a significant blue
shift; if the outflow velocity is tied to the ionizing luminosity, the
low ionizing flux measured in the deep low/hard state observation
could imply an even lower outflow rate since $\dot{M}_{wind} \propto
v$, and a lower kinetic power since $L_{wind} \propto v^{3}$.

If a strong wind did persist in the
low/hard state, it would have to originate very close to the black
hole.  The photoionization models presented in Miller et al.\ (2006)
suggest $\xi = 3-4 \times 10^{5}~ {\rm erg}~ {\rm cm}~ {\rm s}^{-1}$,
but this could potentially be pushed lower, and a reasonable lower
bound would be $\xi \geq 10^{4}$.  For a distance of 8.5~kpc, the flux
measured in the low/hard state corresponds to $L \simeq 2 \times
10^{37}~ {\rm erg}~ {\rm s}^{-1}$.  If the low/hard state wind is at
least as ionized as before, and taking $\xi =L/nr^{2}$, $(2\times
10^{37}~{\rm erg}~ {\rm s}^{-1})/nr^{2} \geq 10^{4}~ {\rm erg}~ {\rm
  cm}~ {\rm s}^{-1}$.  Assuming $N = nr$, then $N \leq (2\times
10^{33}~ {\rm cm}^{-1})/ r$.  If the wind were undiminished, $N =
10^{23}~ {\rm cm}^{-2}$.  This would imply $r = 10^{10}$~cm and $n =
10^{13}~ {\rm cm}^{-3}$.  Such a small radius and high density are
similar to the magnetic wind detected in GRO J1655$-$40 (Miller et
al.\ 2006, 2008, Kallman 2009, also see Neilsen \& Homan 2012).
Moreover, Luketic et al. (2010) have recently shown that winds with $n \geq
10^{12}~ {\rm cm}^{-3}$ are difficult or impossible to drive
thermally.

\section{Discussion and Conclusions}
A dense, equatorial disk wind was previously detected in {\it
  Chandra}/HETGS observations of H~1743$-$322 in disk--dominated
states (Miller et al.\ 2006a).  This source is a strong black hole
candidate based on its X-ray properties (e.g. Homan et al.\ 2005), but
its mass and distance have not been determined.  Assuming fiducial
values of $d = 8.5$~kpc (based on the proximity of the source to the
Galactic center) and $M = 10~M_{\odot}$, the photoionization models
applied in Miller et al. (2006a) describe a wind with $\dot{M}_{out}
=$ 3--4$ \times 10^{17}~ {\rm g}~ {\rm s}^{-1}$ that may originate
within $r \simeq 10^{2-3}~ GM/c^{2}$ of the black hole.

The blue-shifted Fe XXV and Fe XXVI absorption lines found in
the brighter, disk--dominated phases are not detected in this
deep observation.  Importantly, the ionizing photon flux in this
``low/hard'' state was found to be within the range measured when a
disk wind was previously detected.  Using simple arguments, we have
shown that the mass outflow rate is likely reduced by at least a
factor of $\sim$3 in the low/hard state, compared to the strongest
prior line detections, and might be reduced far more severely.  If a
wind with a column density like that measured previously were to
persist in the low/hard state, it would would have to be dense and
originate close to the black hole, and would likely be driven
magnetically (e.g. Miller et al.\ 2008, Luketic et al.\ 2010).

A disk wind was absent in H 1743$-$322 during one of the {\it Chandra}
observations in 2003 (see the right-most limits in Figure 2).  Based
on photoionization modeling, Miller et al.\ (2006a) concluded that the
non-detection likely required a geometric change (lower density,
depth, or covering factor), not merely higher ionization.  Similarly,
Neilsen \& Homan (2012) recently concluded that over-ionization could
not account for variability seen in the disk wind in GRO J1655$-$40.
Stellar-mass black hole winds may be thermally driven: radiation from
the central engine heats gas in the outer disk to the local escape
speed (see, e.g., Begelman, McKee, \& Shields 1983, Woods et
al.\ 1996; also see Luketic et al.\ 2010).  It is not clear that such
winds should be strongly variable.  Geometric changes associated with
state transitions --- such as the presence or absence of an {\it
  inner} disk (e.g., Esin, McClintock, \& Narayan; also see Reis,
Fabian, \& Miller 2010) --- need not affect the {\it outer} disk.  The
important parameter for wind detection is the ionizing photon flux,
which need not change drastically across states.

Ueda et al.\ (2010) suggested a geometric change that
might explain variability in thermal winds: the development of a hot,
geometrically--thick, optically--thick ($\tau =$7--10) corona that can
``shield'' the outer disk.  This
picture may be consistent with the ``very high'' state, and
may be able to account for the prior non-detection of a disk wind
in H~1743$-$322 (Miller et al.\ 2006a).  However, this geometrical
change is not consistent with the ``low/hard'' state, which is
of much greater interest since this is the only state where jets are
produced in a steady fashion (e.g. Fender, Belloni, \& Gallo 2004).
Black hole spectra in the ``low/hard'' state require a relatively low
optical depth and high temperature when fit with Componization models
($\tau \leq$ 1--2, $kT_{e} =$30--120~keV; see e.g. Gierlinski et
al.\ 1997, Torii et al.\ 2011).  Moreover, the outer disk must be
irradiated in order to explain UV and optical emission in the
``low/hard'' state (e.g. Rykoff et al.\ 2007, Reynolds \& Miller
2012).  

Magnetic driving may provide an alternative to thermal driving, and
may be suited to the anti-correlation between winds and jets when
comparing ``high/soft'' and ``low/hard'' states.  For instance, the
magnetic field configuration may change from toroidal to poloidal in
transitions from disk--dominated states to the ``low/hard'' state.
Disk winds might then be driven by magnetic pressure generated in a
thin disk (e.g. Proga 2003, Ohsuga et al.\ 2009) while jets might be
driven by magneto-centrifugal acceleration along poloidal field lines
(Blandford \& Payne 1982), perhaps aided by black hole spin (Blandford
\& Znajek 1977).  This change could be precipitated by a drop in
$\dot{m}$ through the disk; poloidal fields may be easier to anchor in
thicker disks (Reynolds, Garofalo, \& Begelman 2006).  Alternatively,
it is possible that poloidal fields could dominate on each side of a
state transition, and that $\dot{m}_{disk}$ modulates how much mass is
loaded onto poloidal field lines (e.g. Spruit 1996).  Winds would
originate when the mass outflow rate is high, perhaps breaking field
lines or dragging them to make a small angle with respect to the disk
(Proga 2003).  Jets would then originate when the mass outflow rate is
relatively low, consistent with the ``low/hard'' state, allowing for
more effective acceleration.  This latter scenario may be supported by
recent work suggesting that black hole winds and jets may be regulated
in a common fashion across the mass scale (King et al.\ 2012b).

The analysis of wind properties across state transtions presented in
this work may favor a magnetic wind component.  Stronger support for
magnetically--driven winds may derive from photoionization modeling
that is able to infer a small launching radius and/or very high mass
outflow rate (e.g. Miller et al.\ 2008), or perhaps from evidence of
common regulation of the kinetic power in winds and jets (e.g. King et
al.\ 2012b).  To constrain the launching radius, the density of the
gas must be measured directly, since $r^{2} = L / n\xi$.  Currently,
this has only been possible for GRO J1655$-$40, via the detection of
the density-sensitive Fe XXII line pair ($n = 10^{14}~ {\rm
  cm}^{-3}$), and possibly for NGC 4051 (King, Miller, \& Raymond
2012).  In contrast, the density of the disk winds in H 1743$-$322 and
GRS 1915$+$105 has not been directly constrained (but estimates based
on variability are given in Neilsen et al.\ 2011, 2012).  Trends in
these sources may or may not be consistent with thermal driving (Ponti
et al.\ 2012, King et al.\ 2012b).  Observing stellar-mass black holes
with low line-of-sight column densities will facilitate radius and
density constraints through the detection of Fe XXII lines, and help
to better reveal wind launching mechanisms.

\vspace{0.2in}
We thank the anonymous referee.  We thank Harvey Tananbaum for
executing this observation.  ALK acknowledges support
through the NASA Earth and Space Sciences Fellowship.  JMM
acknowledges support through the {\it Chandra} Guest Observer program.



\begin{references}
\reference{} Begelman, M. C., McKee, C. F., \& Shields, 1983, ApJ, 271, 70

\reference{} Belloni, T., Munoz-Darias, T., Motta, S., Stiele, H.,
Carbone, D., 2010, ATEL 2788

\reference{} Blandford, R. D., \& Payne, D. G., 1982, MNRAS, 199, 883

\reference{} Blandford, R. D., \& Znajek, R. L., 1977, MNRAS, 179, 433

\reference{} Blum, J. L., Miller, J. M., Cackett, E., Yamaoka, K.,
Takahashi, H., Raymond, J., Reynolds, C. S., Fabian, A. C., 2010, ApJ,
713, 1244

\reference{} Esin, A. A., McClintock, J. E., \& Narayan, R., 1997, 489, 865

\reference{} Fender, R. P., Belloni, T., \& Gallo, E., 2004, MNRAS, 355, 1105

\reference{} Gierlinksi, M., Zdziarksi, A. A., Done, C., Johnson,
W. N., Ebisawa, K., Ueda, Y., Haardt, F., Philips, B., 1997, MNRAS,
288, 958

\reference{} Homan, J., Miller, J. M., Wijnands, R., van der Klis, M.,
Belloni, T., Steeghs, D., \& Lewin, W. H. G., 2005, ApoJ, 623, 383

\reference{} Kallman, T. R., \& McCray, R., 1982, ApJS, 50, 263

\reference{} Kallman, T. R., Bautista, M. A., Goriely, S., Mendoza,
C., Miller, J. M., Palmeri, P., Qunet, P., Raymond, J., 2009, ApJ,
701, 865

\reference{} King, A. L., Miller, J. M., Cackett, E. M., Fabian,
A. C., Markoff, S., Nowak, M. A., Rupen, M., Gultekin, K., \&
Reynolds, M. T., 2011, ApJ, 729, 19

\reference{} King, A. L., Miller, J. M., \& Raymond, J. C., 2012, ApJ, 746, 2

\reference{} King, A. L., Miller, J. M., Raymond, J., Fabian, A. C.,
Reynolds, C. S., Gultekin, K., Cackett, E. M., Allen, S. W., Proga,
D., Kallman, T. R., 2012, ApJ, subm., arxiv:1205.4222

\reference{} Kubota, A., et al., 2007, PASJ, 59S, 185

\reference{} Luketic, S., Proga, D., Kallman, T. R., Raymond, J. C., Miller, J. M., 2010, ApJ, 719, 515

\reference{} Miller, J. M., et al., 2006a, ApJ, 646, 394

\reference{} Miller, J. M., Raymond, J., Fabian, A., Steeghs, D.,
Homan, J., Reynolds, C., van der Klis, M., Wijnands, R., 2006b, Nature,
441, 953

\reference{} Miller, J. M., Raymond, J., Reynolds, C. S., Fabian,
A. C., Kallman, T. R., \& Homan, J., 2008, ApJ, 680, 1359

\reference{} Neilsen, J., \& Lee, J. C., 2009, Nature, 458, 481

\reference{} Neilsen, J., \& Homan, J., 2012, ApJ, 750, 27

\reference{} Neilsen, J., Remillard, R., \& Lee, J. C., 2011, ApJ, 737, 69

\reference{} Neilsen, J., Remillard, R., \& Lee, J. C., 2012, ApJ, 750, 71

\reference{} Ohsuga, K., Mineshige, S., Mori, M., Kato, Y., 2009, PASJ, 61, 7

\reference{} Ponti, G., Fender, R. P., Begelman, M. C., Dunn, R.,
Neilsen, J., Coriat, M., 2012, MNRAS, 422, L11

\reference{} Proga, D., 2000, ApJ, 538, 684

\reference{} Proga, D., 2003, ApJ, 585, 406

\reference{} Reis, R. C., Fabian, A. C., \& Miller, J. M., 2010 MNRAS, 402, 836

\reference{} Reynolds, C. S., Garofalo, D., Begelman, M. C., 2006, ApJ, 651, 1023

\reference{} Reis, R. C., Fabian, A. C., Miller, J. M., 2010, MNRAS, 402, 836

\reference{} Reynolds, M. T., \& Miller, J. M., 2010, ApJ, 723, 1799

\reference{} Reynolds, M. T., \& Miller, J. M., 2012, ApJ, subm.

\reference{} Russell, D. M., Miller-Jones, J., Maccarone, T., Yang,
Y., Fender, R., Lewis, F., 2011, ApJ, 739, L19

\reference{} Rykoff, E. S., Miller, J. M., Steeghs, D., \& Torres, M. A. P., 2007, ApJ, 666, 1129

\reference{} Spruit, H., 1996, in ``Physical Processes in Binary
Stars'', NATO ASI series, Kluwer Dordrecht, eds. R. A. M. J. Wijers,
M. B. Davies, C. A. Tout, arxiv:astro-ph/9602022

\reference{} Tombesi, F., Cappi, M., Reeves, J. N., Palumbo, G.,
Yaqoob, T., Braito, V., Dadina, M., 2010, A\&A, 521, 57

\reference{} Torii, S., et al., 2011, PASJ, 63S, 771

\reference{} Ueda, Y., Yamaoka, K., \& Remillard, R., 2009, ApJ, 695, 888

\reference{} Ueda, Y., et al., 2010, ApJ, 713, 257

\reference{} Verner, D. A., Verner, E. M., \& Ferland, G. J., 1996, ADNDT, 64, 1





\end{references}
\end{document}